\documentclass[twocolumn]{aastex6}

\begin{document}

\title{Stellar dynamics and star-formation histories of z $\sim$ 1 radio-loud galaxies}
\author{
Ivana Bari\v{s}i\'c$^1$,
Arjen van der Wel$^1$,
Rachel Bezanson$^2$,
Camilla Pacifici$^3$,
Kai Noeske$^4$,
Juan C. Mu\~noz-Mateos$^5$,
Marijn Franx$^6$,
Vernesa Smol\v{c}i\'c$^7$,
Eric F. Bell$^{8}$,
Gabriel Brammer$^9$,
Jo\~{a}o Calhau$^{10}$,
Priscilla Chauk\'{e}$^1$,
Pieter G. van Dokkum$^{11}$,
Josha van Houdt$^1$,
Anna Gallazzi$^{12}$,
Ivo Labb\'e$^6$,
Michael V. Maseda$^6$,
Adam Muzzin$^{13}$,
David Sobral$^{6,10}$,
Caroline Straatman$^1$,
Po-Feng Wu$^1$}


\thanks{$^1$Max-Planck Institut f\"ur Astronomie, K\"onigstuhl 17, D-69117, Heidelberg, Germany}
\thanks{$^2$Department of Astrophysics, Princeton University, Princeton, NJ 08544, USA}
\thanks{$^3$Astrophysics Science Division, Goddard Space Flight Center, Code 665, Greenbelt, MD 20771, USA}
\thanks{$^5$European Southern Observatory, Alonso de C\'{o}rdova 3107, Casilla 19001, Vitacura, Santiago, Chile}
\thanks{$^6$Leiden Observatory, Leiden University, P.O.Box 9513, NL-2300 AA Leiden, The Netherlands}
\thanks{$^7$Department of Physics, Faculty of Science, University of Zagreb,  Bijeni\v{c}ka cesta 32, 10000  Zagreb, Croatia}
\thanks{$^{8}$Department of Astronomy, University of Michigan, 1085 S. University Ave, Ann Arbor, MI 48109, USA}
\thanks{$^9$Space Telescope Science Institute, 3700 San Martin Drive, Baltimore, MD 21218, USA}
\thanks{$^10$Department of Physics, Lancaster University, Lancaster LA1 4YB, UK}
\thanks{$^{11}$Department of Astronomy, Yale University, New Haven, CT06511, USA}
\thanks{$^{12}$INAF-Osservatorio Astrofsico di Arcetri, Largo Enrico Fermi 5, I-50125 Firenze, Italy}
\thanks{$^{13}$Department of Physics and Astronomy, York University, 4700 Keele Street, Toronto, Ontario, ON MJ3 1P3, Canada}

\email{barisic@mpia.de}


\begin{abstract}

We investigate the stellar kinematics and stellar populations of 58 radio-loud galaxies of intermediate luminosities (L$_{3 GHz}$ $>$ 10$^{23}$\,W\,Hz$^{-1}$ ) at 0.6 $\textless$ z $\textless$ 1.
This sample is constructed by cross-matching galaxies from the deep VLT/VIMOS LEGA-C spectroscopic survey with the VLA 3\,GHz dataset. 
The LEGA-C continuum spectra reveal for the first time stellar velocity dispersions and age indicators of z $\sim$ 1 radio galaxies.
We find that $z\sim 1$ radio-loud AGN occur exclusively in predominantly old galaxies with high velocity dispersions: $\sigma_*>$\,175\,km\,s$^{-1}$, corresponding to black hole masses in excess of $10^8$\,M$_{\odot}$.
Furthermore, we confirm that at a fixed stellar mass the fraction of radio-loud AGN at z $\sim$ 1 is 5 - 10 times higher than in the local universe, suggesting that quiescent, massive galaxies at z $\sim$ 1 switch on as radio AGN on average once every Gyr.
Our results strengthen the existing evidence for a link between high black-hole masses, radio loudness and quiescence at z $\sim$ 1.
\end{abstract}

\section{Introduction}

In order to match the stellar and dark matter halo mass functions and reproduce their evolution through cosmic time, semi-analytical and hydrodynamical galaxy formation models rely on two primary feedback channels to decrease the efficiency of star-formation.
These models implement heating by supernovae that lead to a low star-formation efficiency in low-mass dark matter halos \citep{WHITE78,WHITE91,HOPKINS12}. Feedback from super-massive black holes (SMBHs) is implemented to prevent excessive star-formation in high-mass halos \citep{DELUCIA07, CROTON06, BOWER06, VOGELSBERGER14, SCHAYE15}. The physical prescriptions differentiate between radiative-mode feedback and jet-mode feedback. Radiative-mode feedback (‘quasar mode’) is associated with the high accretion rate of the cold gas onto the SMBH and is related to the gas outflows \citep{SHAKURA75, DIMATTEO05}. Jet-mode (‘radio mode’) feedback is associated with a low accretion rate of hot (`coronal') gas onto the SMBH. The feedback loop is thought to exist between the cooling of hot gas that feeds the SMBH \citep[e.g.,][]{BLANTON01} to trigger an active galactic nuclei (AGN) phase that subsequently provides a heating source, counter-acting cooling and preventing further growth in stellar mass. 

Direct observational evidence for a link between AGN and the heating of halo gas is found in massive clusters, where radio jets are seen to produce cavities in the X-ray emitting gas \citep[see][and references therein]{MCNAMARA07,HECKMANN14} and also in early-type galaxies in lower-mass groups where the presence of cold gas and radio jets is linked to the thermodynamical state of the warm/hot gas \citep{WERNER12,WERNER14}.
Furthermore, indirect evidence in the form of a strong correlation between a lack of star formation (quiescence) and the presence of radio AGN has been gathered for galaxies in the local universe \citep[e.g.,][]{MATTHEWS64,KAUFFMANN03,BEST05}. This correlation suggests that massive galaxies spend extended periods in a radio-loud AGN phase, which provides sufficient energy to keep the halo gas from cooling. 

Until recently, radio observations of high-redshift galaxies were limited to the very highest luminosities (L $\textgreater$ 10$^{24}$\,W\,Hz$^{-1}$), where radio AGN hosts are the most extreme galaxies: brightest cluster galaxies, but also star-bursting galaxies \citep{DEBREUCK02,WILLOTT03}.
Deep surveys with the Karl G. Jansky Very Large Array (VLA) are now probing lower luminosities (L $\gtrsim$ 10$^{22}$\,W\,Hz$^{-1}$), enabling us to explore the link between radio AGN and quiescence at large look-back time.
\citet{DONOSO09} showed that the fraction of radio-loud galaxies increases out to z $\sim$ 1, and that its power-law dependence on stellar mass (f$_{radio-loud}$ $\propto$ M$^{2.5}_{*}$) is consistent with what is seen for present-day galaxies \citep{BEST05}. \citet{SIMPSON13} demonstrate that up to z $\sim 1$ radio AGN preferentially reside in galaxies with evolved stellar populations as traced by the D$_n$(4000) index.
\citet{REES16} confirm these results, but show that at $z>1$ radio AGN are hosted more frequently by star-forming galaxies.
Finally, \citet{WILLIAMS15} demonstrate that the fraction of radio-loud AGN of luminosities $\textgreater$ 10$^{24}$\,W\,Hz$^{-1}$ increases out to z = 2, that the power-law mass dependence becomes flatter with the increasing mass, and that the slope of mass dependence becomes shallower with the increasing redshift.

In this study we use deep, rest-frame optical spectra from the Large Early Galaxy Astrophysics Census (LEGA-C) survey of galaxies in the redshift range 0.6 $\textless$ z $\textless$ 1 \citep{WEL16}. The LEGA-C optical spectra provide us for the first time with direct constraints on recent and long-term star-formation histories and stellar dynamical properties of a large sample of galaxies at large look-back time. Cross-matching the LEGA-C sample with the recently completed 3\,GHz VLA survey \citep{SMOLCIC17} allows us to examine for the first time stellar populations and velocity dispersions of intermediate luminosity  radio-loud AGN at these redshifts.  The aim of this paper is to test the hypothesis that radio-loud AGN preferably occur in quiescent galaxies with large velocity dispersions (black hole masses) over a broad range in cosmic time.  The confirmation of this hitherto poorly constrained assumption is crucial for the radio-mode feedback picture.

The outline of this paper is the following. In Section 2 we give an overview of LEGA-C and VLA datasets, and introduce the selection criteria and classification scheme for the radio-loud sub-sample. We present our main results and describe stellar content and star-formation activity of radio-loud AGN in Section 3. A summary of our work is then given in Section 4.

\section{Data, Sample Selection and Classification}
\label{data}
In this section we give an overview of the data sets analysed in this work. We present the criteria adopted for the selection of the radio-loud sub-sample among the whole LEGA-C sample, and we describe the method used to classify the radio-loud galaxies into quiescent and star-forming galaxies. By comparing with local benchmark samples we also measure the evolution of the fraction of radio-loud galaxies out to z $\sim$ 1. 

\subsection{LEGA-C}
The LEGA-C survey \citep{WEL16} is an ESO public spectroscopic survey with VLT/VIMOS \citep{LEFEVRE03} with the aim of obtaining high signal-to-noise ratio (S/N\,$\sim$ 20$\,\rm \AA\,^{-1}$) continuum spectra of 0.6 $\textless$ z $\textless$ 1 galaxies. The full LEGA-C sample will consist of more than 3000 galaxies, $K$-band selected from the \citet{MUZZIN13} UltraVISTA survey in the 1.62 square degree region within the COSMOS field \citep{SCOVILLE07}. The spectral resolution is $R=2500$, spanning the wavelength range from 6300{\,\AA} to 8800{\,\AA}. The current paper uses the Data Release II \footnote{http://www.eso.org/qi/} sample of 1989 galaxies observed during the first two years of LEGA-C observations. This sample is representative of the final sample, which is, in turn, representative of the galaxy population at a given $K$-band flux density. That is, our sample selection is independent of galaxy color and morphology.

In this study we use redshifts, stellar velocity dispersions, D$_n$(4000) break and H$\delta$ absorption indices, nebular emission line equivalent widths, as well as physical parameters estimated from broad-band photometry (UV$+$IR star formation rates (SFR) and stellar masses). UV and IR luminosity based SFRs are estimated following \citet{WHITAKER12b}. Stellar masses are derived using a \citet{CHABRIER03} Initial Mass Function, \citet{CALZETTI00} dust extinction, and using \citet{BRUZUAL03} stellar population libraries and exponentially declining SFR. For further details on the data reduction steps and the method used to derive the physical parameters we refer to \citet{WEL16}.

\subsection{VLA - COSMOS}
We use the observations at 3\,GHz (10 cm) (PI: Vernesa Smol\v ci\' c) covering the 2 square degree COSMOS field, obtained by the VLA radio interferometer. The observations were conducted between 2012 and 2014, with a total observation time of 384 hours, yielding a final mosaic with the angular resolution of 0.75$\,"$ and a median rms of 2.3\,$\mu$Jy\,beam$^{-1}$. For further details we refer to \citet{SMOLCIC17}.

\begin{figure}
\includegraphics[width=\linewidth]{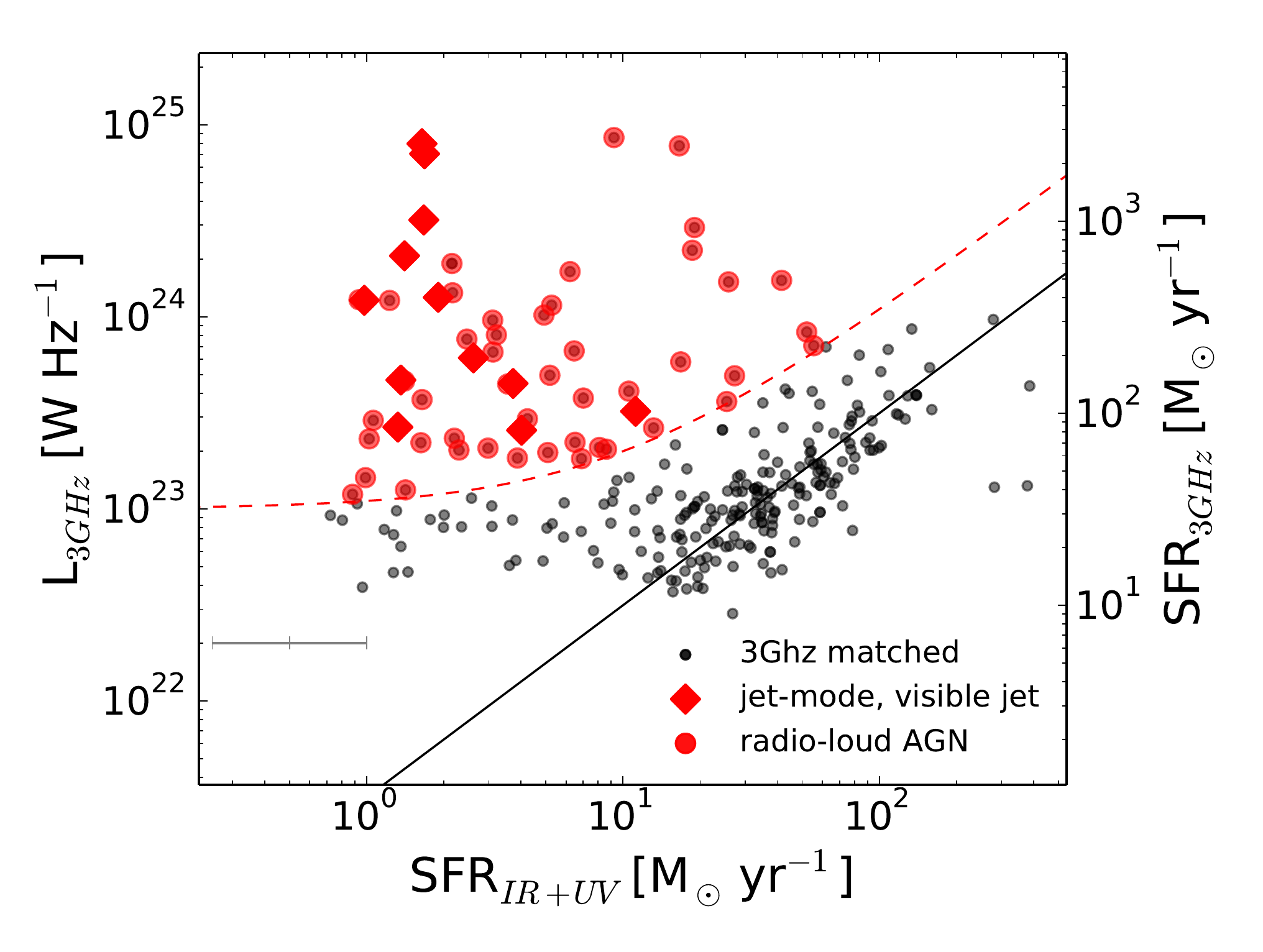}
\caption{The 3\,GHz radio luminosity (left-hand y-axis) and implied SFR (right-hand y-axis) versus SFR$_{UV+IR}$ for the LEGA-C + VLA cross-matched sample. Red points meet our luminosity criteria for radio-loud AGN, red diamonds have visible jets in the radio image. Grey error bar represents the typical uncertainty of the SFR$_{UV+IR}$. The 3\,GHz luminosity uncertainties for radio-loud AGN are smaller than the size of the data points.}
\label{lum}
\end{figure}

\subsection{Selection of radio-loud AGN}
\label{method}

The current LEGA-C dataset contains 1989 galaxies, out of which 322 are found to have a radio counterpart after cross-matching the LEGA-C survey catalog and the VLA 3\,GHz 5.5\,$\sigma $ catalog \citep{SMOLCIC17}. 
The matching radius between LEGA-C and VLA coordinates is 0.7$\,"$. The systematic offset in right ascension and declination is 0.1$\,"$ and 0.03$\,"$ respectively, while the random offset is 0.1$\,"$.
We convert radio continuum fluxes for cross-matched sources to luminosities, using VLA 3\,GHz flux densities and LEGA-C spectroscopic redshifts, following the \citet{CONDON92} luminosity relation. Implied radio-based star formation rates are then found from luminosities using \citet{BELL03} calibration of the radio-FIR correlation.


\begin{figure}
\includegraphics[width = \linewidth]{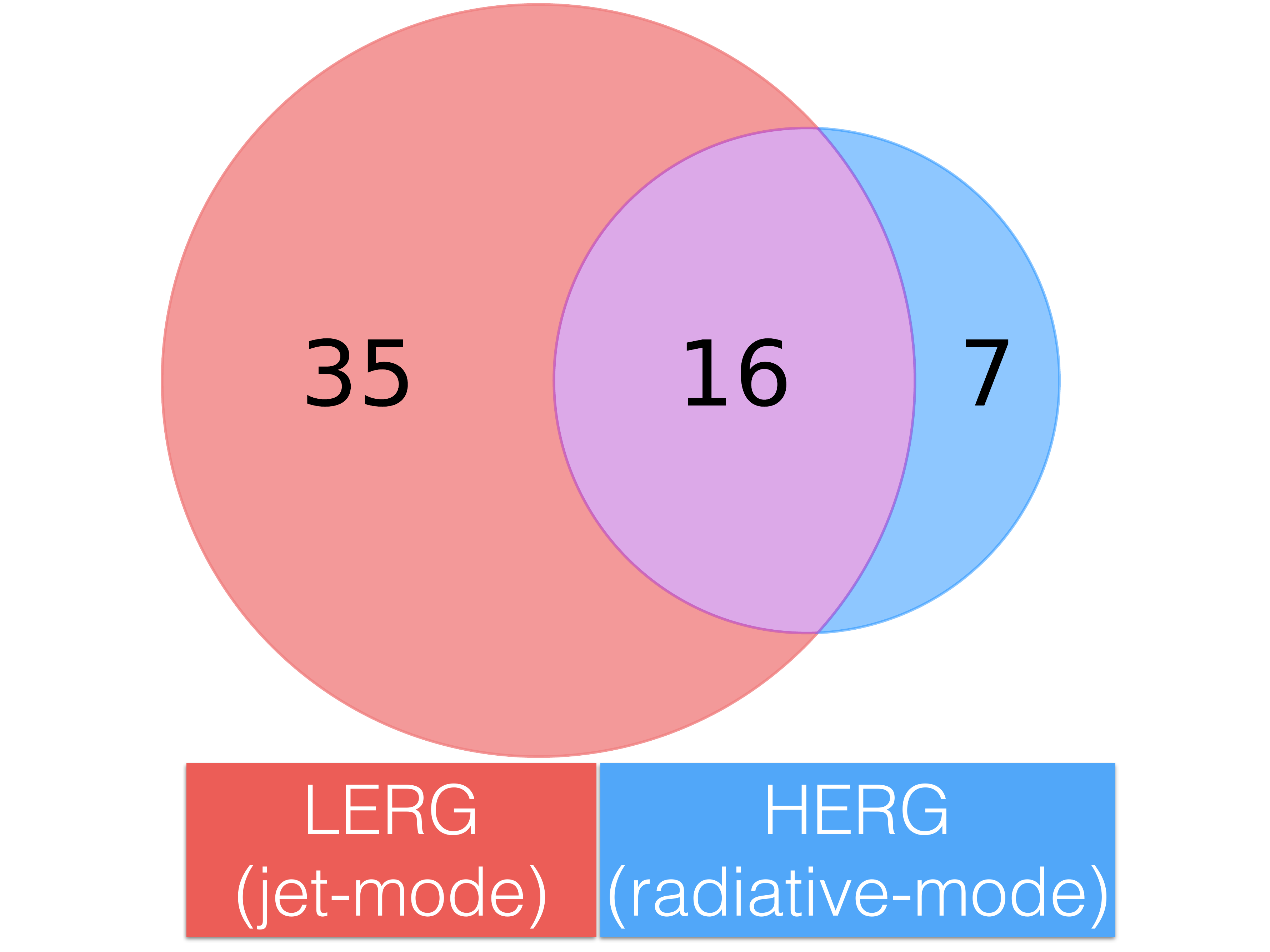}
\caption{Classification of galaxies on jet-mode / radiative-mode based on the ratio of optical emission lines [OIII]/H$\beta$ and [OII]/H$\beta$ (see Section~\ref{clas} for details).}
\label{venn}
\end{figure}

\begin{figure*}
\includegraphics[height = 9in]{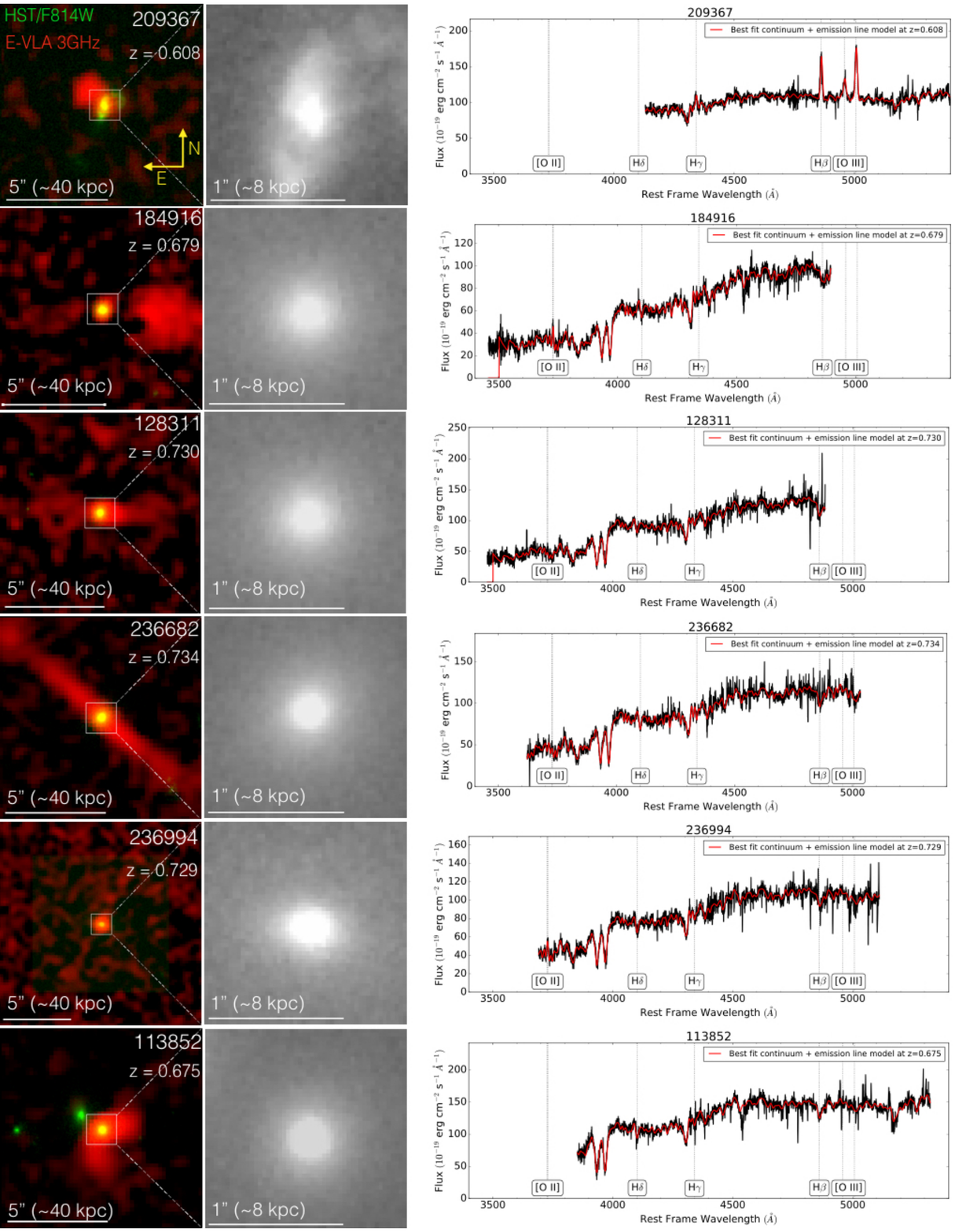}
\caption{False-color HST/ACS/F814W+VLA/3\,GHz images of radio sources with visible jets in the 3\,GHz data, along with the corresponding LEGA-C optical spectra (black) with the best-fitting stellar continuum model (red). }
\label{cut}
\end{figure*}

\renewcommand{\thefigure}{\arabic{figure} (Cont.)}
\addtocounter{figure}{-1}

\begin{figure*}
\includegraphics[height = 9in]{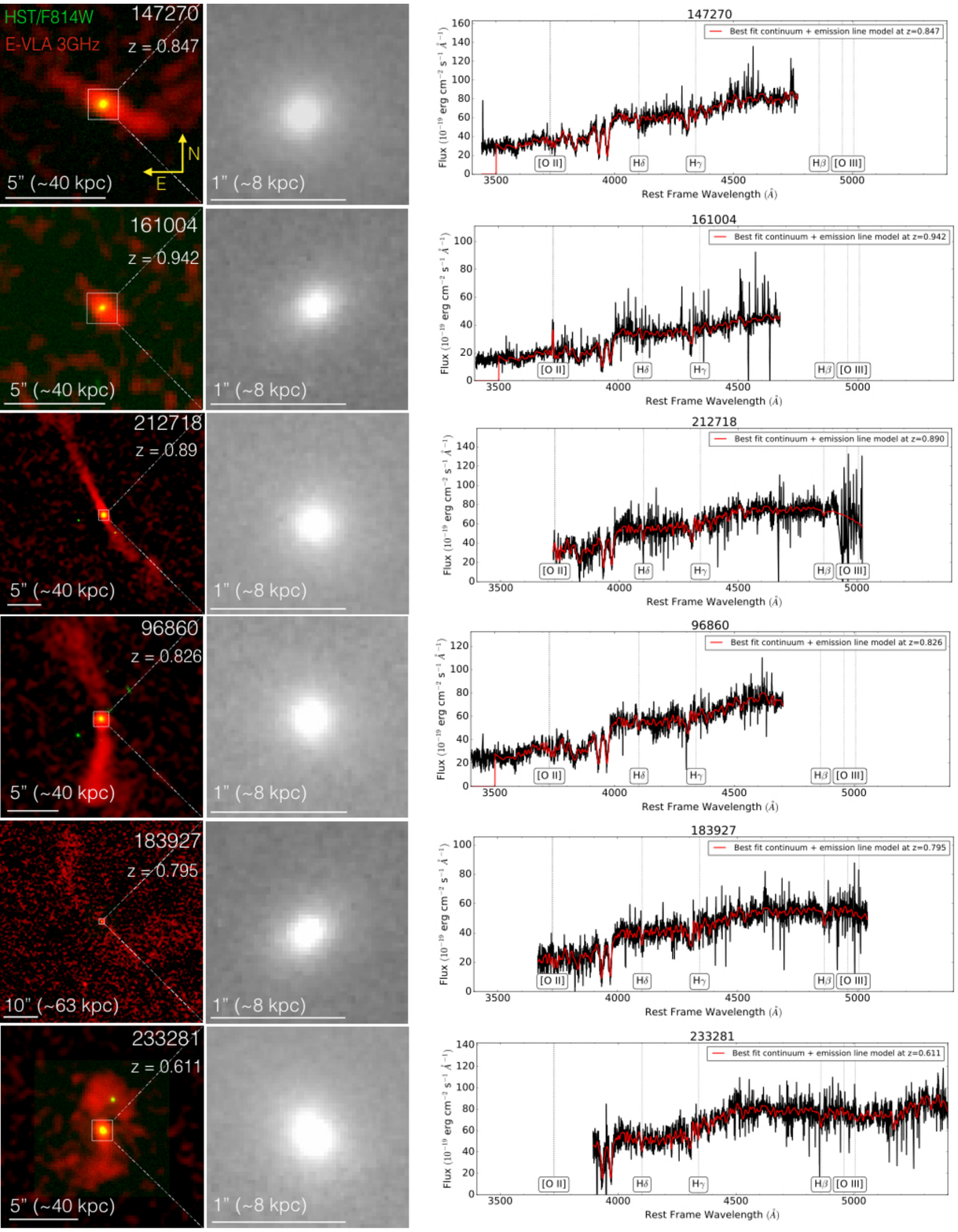}
\caption{False-color HST/ACS/F814W+VLA/3\,GHz images of radio sources with visible jets in the 3\,GHz data, along with the corresponding LEGA-C optical spectra (black) with the best-fitting stellar continuum model (red).}
\end{figure*}

\renewcommand{\thefigure}{\arabic{figure}}

Figure \ref{lum} compares the radio luminosity, L$_{3GHz}$ (and the corresponding star-formation rate) with the UV$+$IR-based SFR for this cross-matched sample. The good correspondence between the star-formation rate indicators is by construction, as the radio-based SFR is calibrated with the SFR derived from the IR luminosity. 
We see no evidence for a change in 
the far-infrared to radio luminosity ratio $q$, but this is not inconsistent with the evidence for such evolution \citep{MAGNELLI15, DELHAIZE17} obtained from larger samples across a much larger redshift range than $z=0$ to $z=1$.

Of interest for this study are the sources that are outliers from the one-to-one relation (black line), that is, the sources with excessive radio luminosities. We estimate the radio AGN luminosity by subtracting $3\times$ the radio luminosity expected on the basis of the UV$+$IR-based SFR from the observed radio luminosity. That is, we allow for a factor 3 scatter in the radio-SFR relation. Adopting the radio luminosity limit from \cite{BEST05}, we select those 58 galaxies with radio AGN luminosities $>10^{23}$\,W\,Hz$^{-1}$  as our radio-loud AGN sample (see Figure \ref{lum}, red dashed curve).



\subsection{Classification of the radio-loud objects}
\label{clas}


We examine the optical spectra of the 58 radio-loud AGN in order to distinguish between jet-mode (low-excitation radio galaxy, LERG) and radiative-mode (high-excitation radio galaxy, HERG) based on the presence or absence of strong high-excitation emission lines \citep{HINE79, LAING94}. 
We classify a radio galaxy as a LERG if there are no or only low-excitation (Balmer) emission lines (EW([OIII], [OII], [H$\beta$])\,$\textgreater$\,-5\,\AA). For systems with strong emission lines we classify those with high [OIII]$\lambda$5007$/$H$\beta$ and/or [OII]$\lambda$3727$/$H$\beta$ ratios ($\textgreater$ 1) as HERGs.
In our sample of 58 radio-loud AGN, 35 are classified as LERGs and 7 as HERG with high confidence. The classification of 16 objects remains undetermined, as we were not able to discriminate with high confidence between jet/radiative-mode. For the remainder of this paper we consider these objects as LERG with 50\% probability for the purpose of statistical calculations.
In Figure \ref{venn} we summarize the results of our classification, which illustrates that almost 60\% of our sample are LERGs. We list the properties of the 58 radio AGN in Table~\ref{table}. 



We examine the VLA 3\,GHz images of our sample for visual confirmation of radio jets. Forty six radio-loud galaxies are consistent with point sources, but in 12 cases we identify extended morphologies. We show the VLA 3\,GHz and HST F814W images as well as the LEGA-C spectra of these 12 objects in Figure \ref{cut}.  

\section{Properties of $z\sim1$ jet-mode galaxies}

\subsection{Fraction of jet-mode galaxies}
\begin{figure}
\includegraphics[height = 2.5in]{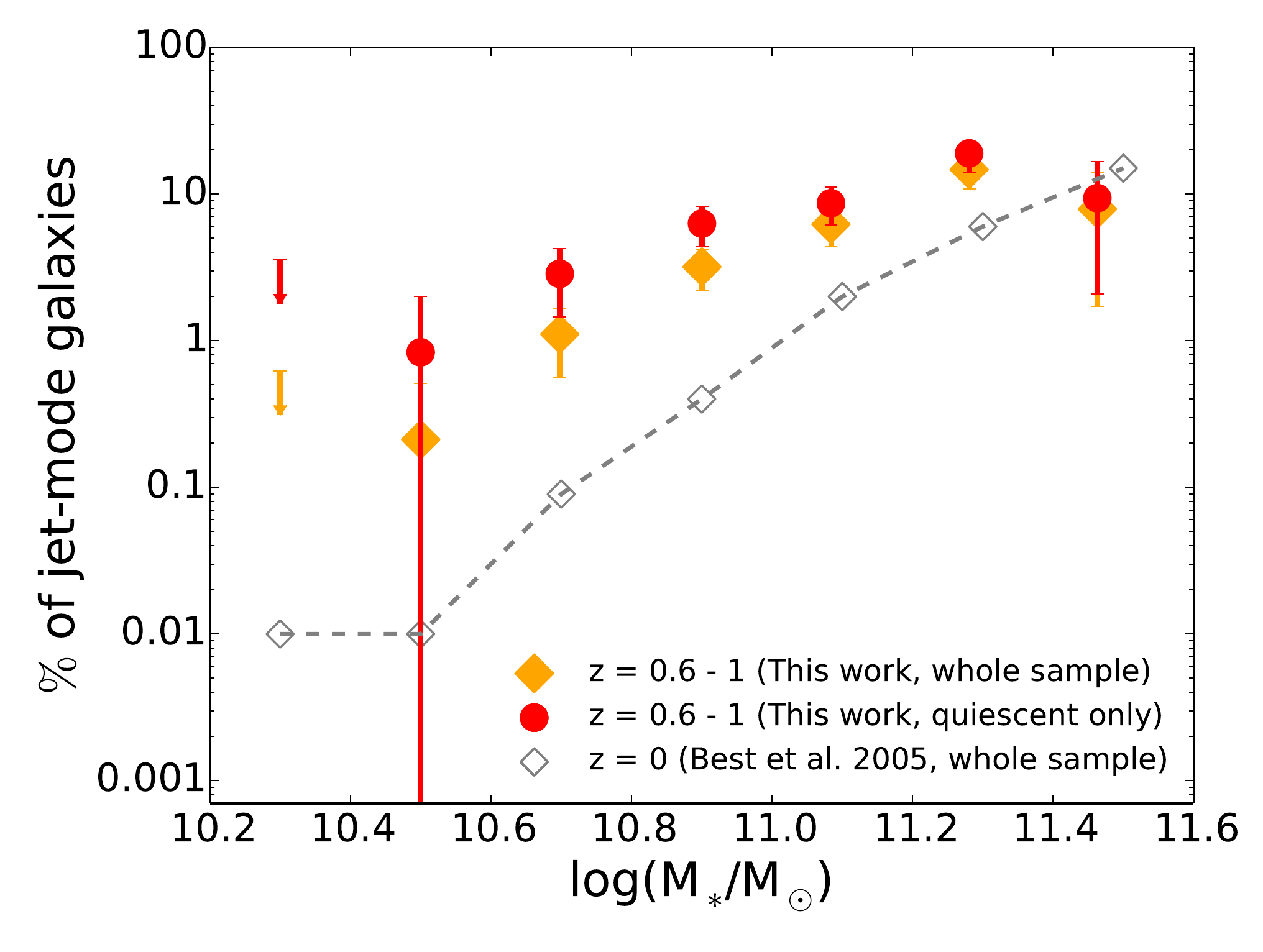}
\caption{Fraction of jet-mode galaxies among all and quiescent galaxies in the LEGA-C as a function of the stellar mass (orange diamonds and red circles respectively). Open orange diamonds show the fraction of jet-mode galaxies for the sample of present-day galaxies.}
\label{q_ws}
\end{figure}

Using the selection criteria defined in Section \ref{clas}, we determine the fraction of jet-mode radio galaxies in the LEGA-C sample, considering both star-forming and quiescent galaxies (see Figure \ref{q_ws}).  Since we adopt the radio-loud AGN selection criteria from \citet{BEST05}, we compare our fraction of jet-mode galaxies at a redshift range 0.6 $\textless$ z $\textless$ 1  to their fraction of jet-mode galaxies for present-day galaxies. At fixed mass the fraction of jet-mode galaxies is on average 5\,-\,10 times higher at 0.6 $\textless$ z $\textless$ 1 compared to the present-day universe. Furthermore, we notice the flattening of the power-law mass dependence for the highest mass bin. These findings are consistent with the measurements from \citet{DONOSO09}.


\begin{deluxetable*}{ccccccccc}
\tabletypesize{\scriptsize}
\tablecaption{Physical properties of the observed sub-sample of radio-loud galaxies \label{table}}
\tablewidth{0pt}
\tablehead{
\colhead{ID}     & \colhead{R.A.}         & \colhead{Dec.}      & \colhead{z$_{spec}$} & \colhead{L $_{3GHz}$ {[}W\, Hz$^{-1}${]}$\cdot$10$^{23}$ $^{}$} & \colhead{$\alpha^{1.4}_{3}$ $^a$} & \colhead{d {[}kpc{]} $^{b}$} & \colhead{flag ${^c}$}} 
\startdata
126578 & 150.09805 & 2.281367 & 0.750 & 2.89 $\pm$ 0.18 & -0.46 $\pm$ 0.14 & \nodata & 1\\
140050 & 150.14168 & 2.446062 & 0.899 & 8.04 $\pm$ 0.42 & -0.03 $\pm$ 0.11 & \nodata & 0.5 \\
209637 & 150.10378 & 2.507702 & 0.608 & 3.21 $\pm$ 0.17 & -0.93 $\pm$ 0.19 & 17.69 $\pm$ 5.10 & 0 \\
131063 & 150.35016 & 2.334642 & 0.667 & 12.27 $\pm$ 0.63 & -0.41 $\pm$ 0.09 & \nodata & 1 \\
182797 & 150.39577 & 2.481532 & 0.903 & 2.22 $\pm$ 0.17 & $\textless$ -0.42 
& \nodata & 1 \\
184916 & 150.42589 & 2.513884 & 0.679 & 6.11 $\pm$ 0.33 & $\textless$ 1.77 
& 50.00 $\pm$ 5.36 & 1 \\
185625 & 150.42506 & 2.524558 & 0.984 & 1.83 $\pm$ 0.18 & $\textless$ -0.89 
& \nodata & 0.5 \\
210716 & 150.17261 & 2.523343 & 0.69 & 12.21 $\pm$ 0.60 & -0.80 $\pm$ 0.08 & \nodata & 1 \\
106926 & 150.29488 & 2.034494 & 0.955 & 77.68 $\pm$ 3.91 &  1.02 $\pm$ 0.09 & \nodata & 0.5 \\
108227 & 150.26643 & 2.049850 & 0.960 & 7.07 $\pm$ 0.39 & -0.41 $\pm$ 0.31 & \nodata & 0.5 \\
109352 & 150.10892 & 2.063952 & 0.724 & 4.96 $\pm$ 0.28 & $-0.49^{+0.16}_{-0.17}$ & \nodata & 1 \\
110509 & 150.26849 & 2.077003 & 0.667 & 6.64 $\pm$ 0.34 & -0.06 $\pm$ 0.14 & \nodata & 1 \\
110805 & 150.17149 & 2.084074 & 0.729 & 3.77 $\pm$ 0.20 & $-0.82^{+0.16}_{-0.17}$ & \nodata & 0.5 \\
113394 & 150.35657 & 2.117532 & 0.875 & 2.02 $\pm$ 0.16 & $\textless$ -0.48 
& \nodata & 1 \\
128311 & 150.05669 & 2.301382 & 0.730 & 4.49 $\pm$ 0.24 & -0.50 $\pm$ 0.28 & 19.12 $\pm$ 5.52 & 1 \\
129746 & 150.02608 & 2.318864 & 0.941 & 3.40 $\pm$ 0.24 & $\textless$ 0.13 
& \nodata & 0.5 \\
205180 & 150.00731 & 2.453467 & 0.730 & 18.94 $\pm$ 0.96 & $-0.43^{+0.07}_{-0.08}$ & \nodata & 1 \\
209377 & 150.02267 & 2.508070 & 0.746 & 85.76 $\pm$ 4.47 & -0.79 $\pm$ 0.07 & \nodata & 0 \\
210031 & 150.02242 & 2.516584 & 0.679 & 5.82 $\pm$ 0.29 & -0.93 $\pm$ 0.13 & \nodata & 0 \\
210739 & 150.00941 & 2.526713 & 0.733 & 4.09 $\pm$ 0.24 & -0.49 $\pm$ 0.25 & \nodata & 1 \\
234067 & 149.85017 & 2.452237 & 0.714 & 29.15 $\pm$ 1.48 & -0.35 $\pm$ 0.07 & \nodata & 1 \\
236682 & 149.87180 & 2.479084 & 0.734 & 20.77 $\pm$ 1.04 & -0.78 $\pm$ 0.07 & 95.81 $\pm$ 5.53 & 1 \\
236994 & 149.86151 & 2.484360 & 0.730 & 2.57 $\pm$ 0.16 & -0.65 $\pm$ 0.29 & 109.57 $\pm$ 5.52 & 1 \\
129631 & 149.98328 & 2.317157 & 0.934 & 9.60 $\pm$ 0.49 & -0.61 $\pm$ 0.16 & \nodata & 0.5 \\
131657 & 149.95264 & 2.341849 & 0.945 & 2.33 $\pm$ 0.19 & -1.06 $\pm$ 0.20 & \nodata & 1 \\
169076 & 149.78040 & 2.318275 & 0.677 & 1.96 $\pm$ 0.12 & -0.43 $\pm$ 0.17 & \nodata & 1 \\
169901 & 149.79379 & 2.327209 & 0.893 & 2.64 $\pm$ 0.19 & $\textless$ -0.23 
& \nodata & 0 \\
210564 & 149.91573 & 2.521326 & 0.729 & 6.55 $\pm$ 0.35 & -0.47 $\pm$ 0.14 & \nodata & 1 \\
235394 & 149.76112 & 2.460729 & 0.671 & 4.92 $\pm$ 0.25 & $0.09^{+0.28}_{-0.29}$ & \nodata & 1 \\
235431 & 149.78880 & 2.466439 & 0.732 & 1.19 $\pm$ 0.11 & $\textless$ -0.61 
& \nodata & 0.5 \\
237437 & 149.79221 & 2.489063 & 0.734 & 1.25 $\pm$ 0.11 & $\textless$ -0.56 
& \nodata & 1 \\
111543 & 149.91492 & 2.094372 & 0.884 & 2.09 $\pm$ 0.17 & $\textless$-0.54 & \nodata & 0.5 \\
113852 & 150.01424 & 2.123182 & 0.675 & 31.92 $\pm$ 1.59 & -0.47 $\pm$ 0.07 & 35.63 $\pm$ 5.34 & 1 \\
125257 & 150.06847 & 2.265479 & 0.979 & 3.62 $\pm$ 0.25 & $\textless$ -0.04 
& \nodata & 0.5 \\
147270 & 149.87502 & 2.062635 & 0.847 & 12.63 $\pm$ 0.65 & -1.28 $\pm$ 0.07 & 51.16 $\pm$ 5.81 & 1 \\
151161 & 149.89481 & 2.109374 & 0.666 & 7.65 $\pm$ 0.43 & -0.004 $\pm$ 0.14 & \nodata & 1 \\
161004 & 149.83919 & 2.226176 & 0.943 & 4.68 $\pm$ 0.28 & -0.82 $\pm$ 0.32 & 24.00 $\pm$ 6.00 & 0.5 \\
105328 & 149.90935 & 2.013062 & 0.848  & 1.84 $\pm$ 0.15 & $\textless$ -0.45 & \nodata &  1 \\ 
117992 & 149.94199 & 2.173145 & 0.688  & 2.31 $\pm$ 0.14 & -0.28 $\pm$ 0.32 & \nodata & 1 \\
120120 & 149.99265 & 2.202235 & 0.629  & 2.21 $\pm$ 0.13 & -0.59$^{+0.27}_{-0.25}$ & \nodata & 1 \\
157229 & 149.74300 & 2.179562 & 0.631  & 15.21 $\pm$ 0.75 & -0.64 $\pm$ 0.08 & \nodata & 0.5 \\
212718 & 150.07712 & 2.548955 & 0.890  & 70.58 $\pm$ 0.62 & $\textless$ -1.03 & 247.90 $\pm$ 5.84 & 1 \\
203666 & 150.39935 & 2.794159 & 0.822  & 8.33 $\pm$ 0.43 & -0.95 $\pm$ 0.19 & \nodata &  0.5 \\
215835 & 150.24612 & 2.585822 & 0.675  & 10.21 $\pm$ 0.53 & 0.34 $\pm$ 0.15 &   & 0.5 \\
217020 & 150.16193 & 2.601267 & 0.893  & 2.07 $\pm$ 0.18 & $\textless$ -0.55 & \nodata & 1 \\
218725 & 150.04684 & 2.620396 & 0.736  & 2.94 $\pm$ 0.18 & -1.03 $\pm$ 0.11 & \nodata & 0 \\
232020 & 150.01646 & 2.784381 & 0.983  & 15.48 $\pm$ 0.78 & -0.69 $\pm$ 0.17 & \nodata & 0.5 \\
232196 & 149.98419 & 2.787762 & 0.853  & 4.63 $\pm$ 0.28 & -0.96 $\pm$ 0.33 & \nodata & 0.5 \\
245325 & 149.88518 & 2.581121 & 0.694  & 4.47 $\pm$ 0.24 & -0.13 $\pm$ 0.11 & \nodata & 1 \\
94215 & 150.68156 & 2.324819 & 0.978  & 2.04 $\pm$ 0.20 & $\textless$ -0.81 & \nodata & 0 \\
94982 & 150.63631 & 2.333361 & 0.609  & 11.47 $\pm$ 0.63 & -0.32 $\pm$ 0.09 & \nodata & 1 \\
96860 & 150.66121 & 2.364529 & 0.826  & 79.72 $\pm$ 0.52 & $\textless$ -1.03 & 150.99 $\pm$ 5.77 & 1 \\
182890 & 150.61380 & 2.484840 & 0.744  & 1.45 $\pm$ 0.11 & $\textless$ -0.41 & \nodata & 1 \\
183927 & 150.61508 & 2.500369 & 0.796  & 2.66 $\pm$ 0.17 & $\textless$ 0.26 & 458.54 $\pm$ 5.69 & 1 \\
225672 & 149.91795 & 2.701692 & 0.892  & 17.19 $\pm$ 0.85 & -0.75 $\pm$ 0.1 & \nodata & 0 \\
233281 & 149.94615 & 2.801806 & 0.611  & 12.19 $\pm$ 0.61 & -1.20 $\pm$ 0.23 & 59.58 $\pm$ 5.11 & 1 \\
250117 & 149.77776 & 2.645909 & 0.737  & 22.20 $\pm$  1.12  & -0.52 $\pm$ 0.08 & \nodata & 0.5 \\ 
27265 & 150.14487 & 1.776603 & 0.733  & 13.33 $\pm$ 0.39 & $\textless$ 2.21 & \nodata & 1 \\ \hline
\enddata
\tablenotetext{a}{Radio spectral slope $\alpha$ inferred using S $\propto$ $\nu ^{\alpha}$ at 1.4\,GHz and 3\,GHz}
\tablenotetext{b}{Linear size (diameter) of the jet, with the errors estimated from the beam size (0.75")}
\tablenotetext{c}{Classification of galaxies onto LERG (1) or HERG (0). For conflicting indicators, we classify galaxies as 50\% LERG (0.5)}
\end{deluxetable*}
\newpage

\begin{figure*}
\includegraphics[width = 0.5 \textwidth]{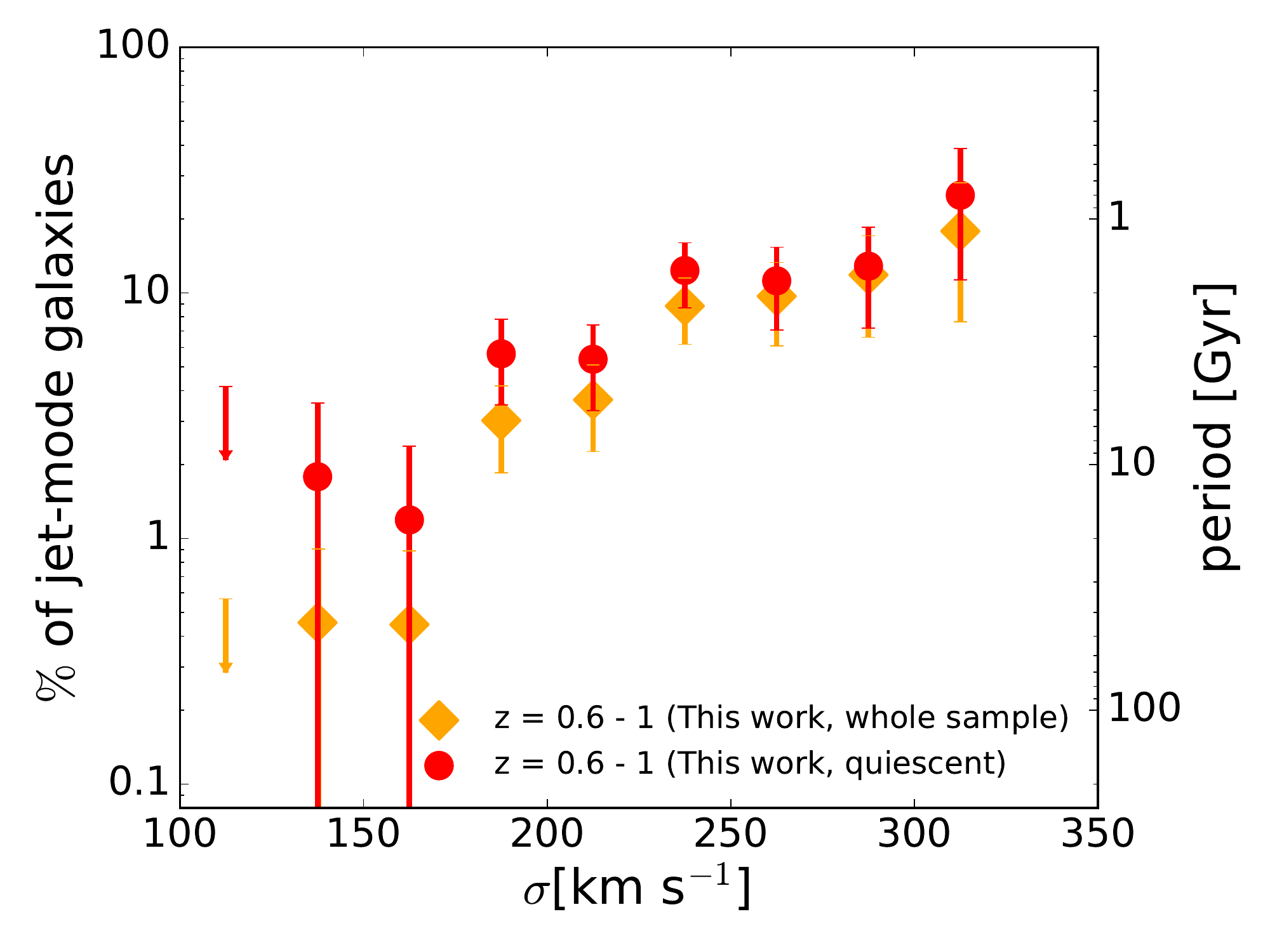}
\includegraphics[width=0.5\textwidth]{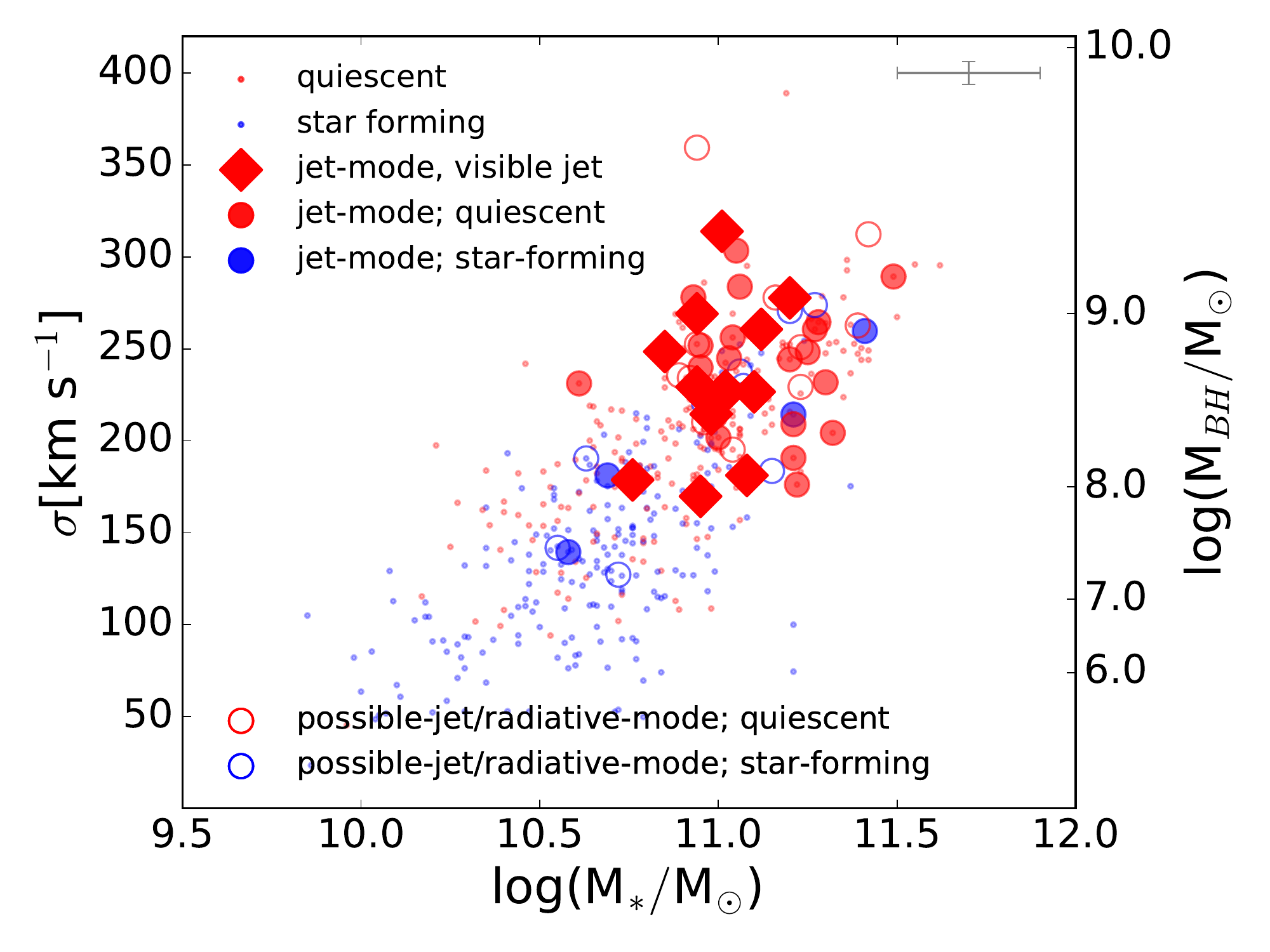}
\caption{\textbf{Left}: Fraction of jet-mode galaxies among the full LEGA-C sample (orange diamonds) and among the quiescent galaxies the LEGA-C sample (red diamonds) as a function of the stellar velocity dispersion $\sigma_* $, \textbf{Right}: Stellar velocity dispersion $\sigma_* $ as a function of the stellar mass. Blue and red points mark star-forming and quiescent galaxies, respectively. Large symbols represent radio AGN, with filled symbols showing those that are securely classified as jet-mode / low-excitation AGN (see Section 2.4). Large diamonds represent those with visible jets in the radio image (see Figure 3).
 Jet-mode AGN occur in massive galaxies with high stellar velocity dispersions, and old stellar populations. Typical error bars are indicated in the corners.}
\label{a}
\end{figure*}


One possible evolutionary scenario is that galaxies grow in stellar mass by a factor of two from z $\sim$ 1 to the present while conserving the fraction of jet-mode galaxies. 
This would imply that growth in stellar mass exceeds growth in black hole mass if we assume that the black hole mass is the only factor that sets the probability or frequency of becoming a radio galaxy. Alternatively, if BHs and stellar mass grow in lockstep \citep[e.g.,][]{CALHAU17}, then more frequent (or longer) radio-loud AGN phases at fixed BH mass could be understood by shorter cooling times at earlier cosmic times. 
The evolution in the fraction of jet-mode AGN in quiescent galaxies must be even faster than that for the general population, given that the fraction of quiescent galaxies is lower at $z\sim1$ than at the present day \citep[e.g.][]{BELL04, FABER07}. We show the fraction of jet-mode AGN in quiescent galaxies in Figure~\ref{q_ws}, and see that above $10^{11}$\,M$_{\odot}$ the fraction of jet-mode galaxies reaches 20\,\%.

\begin{figure}[h!]
\includegraphics[height = 2.5in]{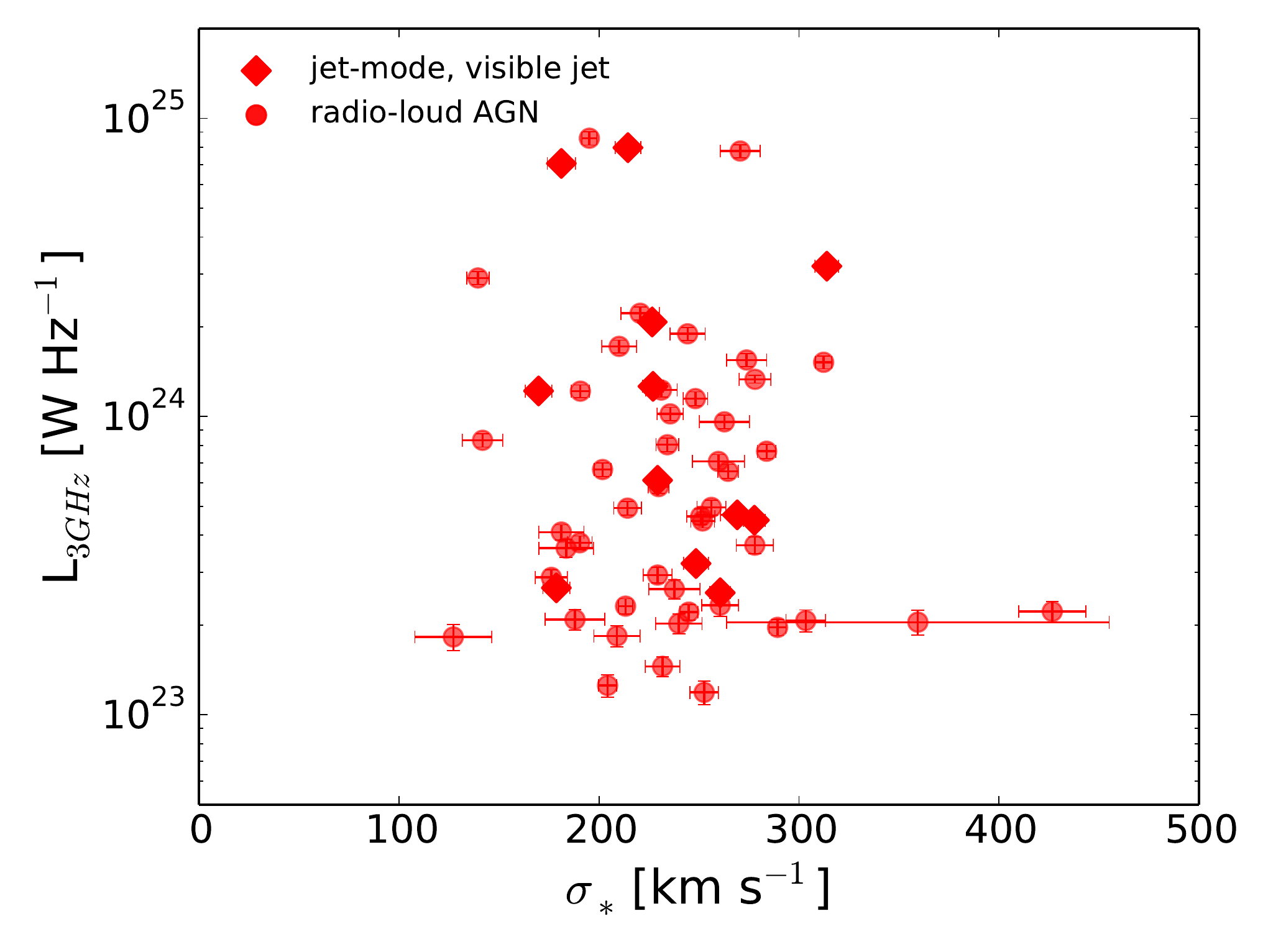}
\caption{The 3 GHz luminosity as a function of the stellar velocity dispersion $\sigma_*$ for the LEGA-C + VLA cross-matched sample of galaxies.}
\label{lum2}
\end{figure}

The fraction of jet-mode AGN also strongly depends on stellar velocity dispersion $\sigma_*$, which can be seen as a proxy for black hole mass: in Figure~\ref{a} we show that more than $\sim$10\% of galaxies with $\sigma_* \gtrsim$\,200\,km\,s$^{-1}$ have jet-mode AGN, reaching 20\,-\,30\% at $\sigma_* \sim$\,300\,km\,s$^{-1}$.
This behavior is only weakly dependent on star-formation activity if at all, which suggests that jet-mode AGN are not associated with quiescence but with high $\sigma_* $. Below $\sigma_*$ =\,175\,km\,s$^{-1}$ we find three jet-mode AGN (one 100\% and two 50\% objects), whereas if the power-law trend seen at high $\sigma_* $ were to continue to low $\sigma_* $ we would expect a total of ten jet-mode AGN in the three low-$\sigma_* $ bins. The detection of only three jet-mode AGN may suggest a threshold black-hole mass of $\sim 10^8$\,M$_{\odot}$ for jet-mode AGN, as inferred from the local black-hole mass\,-\,$\sigma_* $ relation \citep{GEBHARDT03, BEIFIORI12, BOSCH16}. We argue that this threshold is not artificially introduced by our radio luminosity limit, as there is virtually no correlation between $\sigma_* $ and radio luminosity, as shown in Figure \ref{lum2}.

Assuming that the jet-mode fraction of $10-30\%$ at high $\sigma_* $ can be interpreted as a duty cycle, we convert these fractions into the period (or frequency) at which galaxies turn on a jet-mode AGN. In order to make the conversion of the fraction of jet-mode AGN into the period, we need a life time of an AGN jet. Examination of the jet structure morphology of the 12 objects in Figure \ref{cut} reveals that they are reminiscent of the classical Fanaroff-Riley I (FRI) type radio galaxy \citep{FANAROFF74, LEDLOW96}. It has been argued that life times evolve with redshift \citep{ATHREYA98}, but, among our radio galaxies that are detected in the pre-existing 1.4\,GHz VLA data \citep{SCHINNERER10} we find that the spectral slopes, and therefore presumably the ages, are similar to local counterparts  (see Table~\ref{table}). The typical spectral slopes of local counterparts range between -1.3 $\textless$ $\alpha$ $\textless$ -0.5 with the average spectral slope being -0.8 \citep{CONDON92}. \citet{PARMA98} find a correlation for FRI radio galaxies between the linear size of the jet and the synchrotron age of the jet as traced by the spectral slope, implying typical jet ages of about 100 Myr with an uncertainty of at most a factor 2. We therefore assume a lifetime of $2\times100$\,Myr and show the resulting periods in Figure~\ref{a} on the left. We conclude that galaxies turn on a jet-mode AGN about once every Gyr provided that it has a stellar velocity dispersion in excess of $\sigma_*$ =\,175\,km\,s$^{-1}$, corresponding with a black hole mass of 10$^8$\,M$_{\odot}$. Remarkably, this is the same black-hole mass threshold that has been shown to separate quiescent and star-forming galaxies in the local universe \citep{TERRAZAS16}.

\subsection{Stellar Populations of Galaxies with Radio-Loud AGN} 
\label{results}

\begin{figure}[h!]
\includegraphics[width = \linewidth]{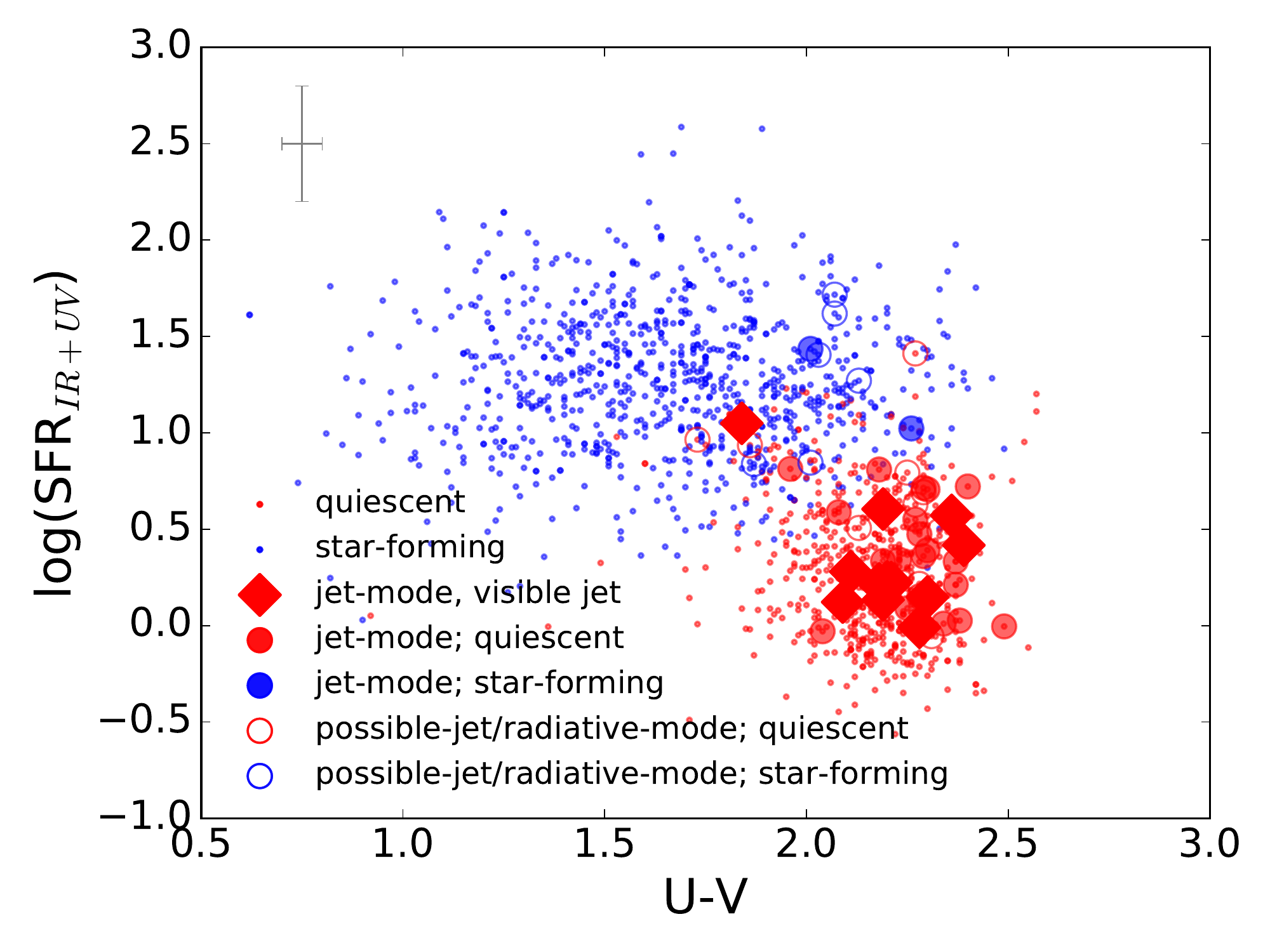}
\caption{SFR$_{UV+IR}$ as a function of the rest-frame U-V color. The symbols are described in Figure 5.  Jet-mode AGN reside in red, quiescent galaxies as had been demonstrated before.}
\label{uv}
\end{figure}

Our $z\sim 1$ radio AGN typically live in red, quiescent galaxies (Figure \ref{uv}) as was shown before by \cite{SMOLCIC09}, \cite{SIMPSON13}, and \cite{REES16}.  

\begin{figure}
\includegraphics[width=0.5\textwidth]{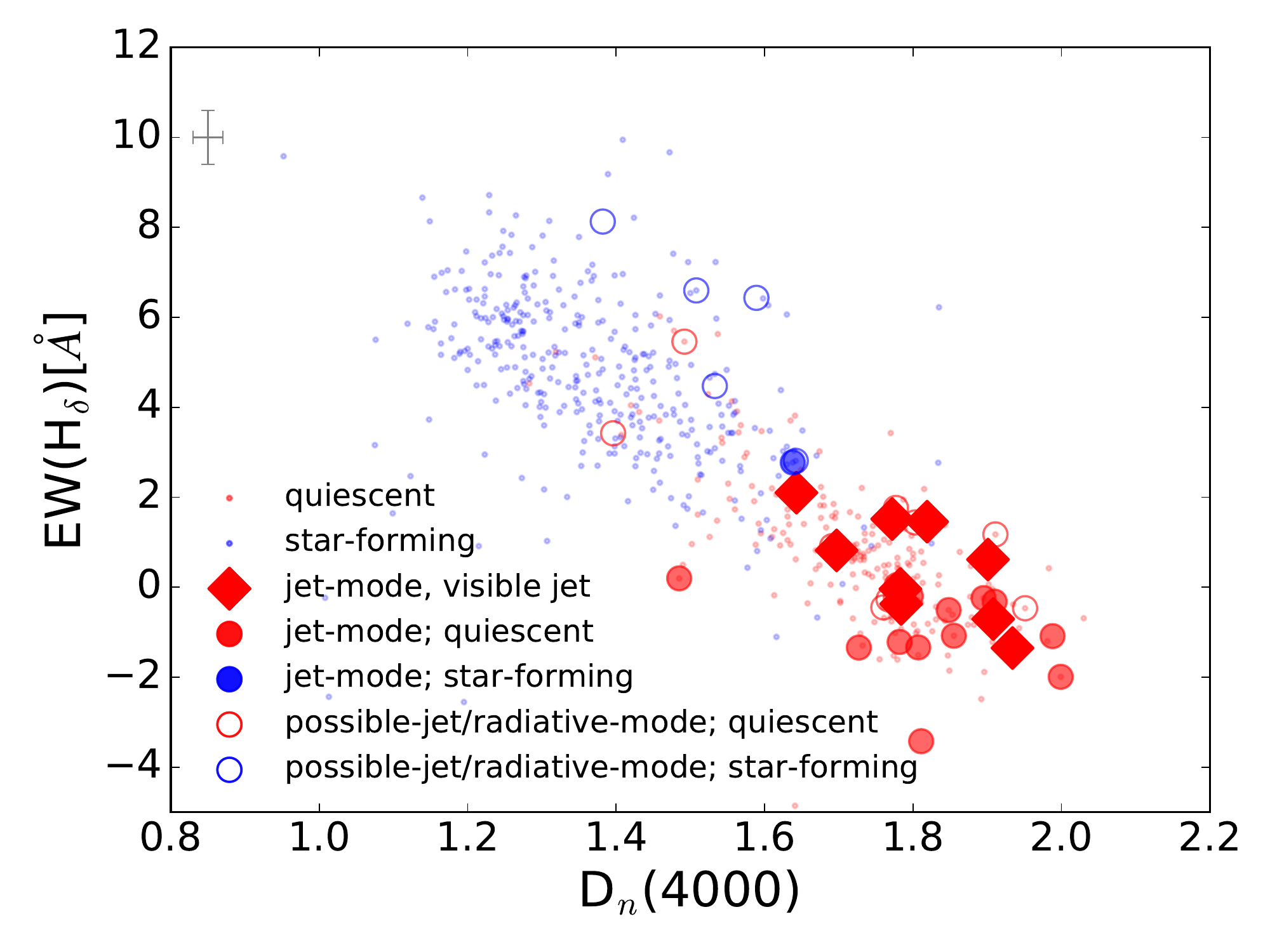}
\caption{Balmer absorption line index H$\delta$ as a function of the 4000\AA\ break D$_n$(4000). The symbols are explained in Figure 5. Jet-mode AGN occur in galaxies with old stellar populations, with little star-formation activity for at least a Gyr.}
\label{aa}
\end{figure}


With the LEGA-C spectra we can examine, for the first time, the detailed stellar population properties of $z\sim1$ radio galaxies. Figure \ref{aa} shows the Balmer absorption line index H$\delta$ as a function of the 4000\AA\ break D$_n$(4000) \citep[][Wu et al., in prep]{BRUZUAL83, BALOGH99, KAUFFMANN03_2}. Both of these parameters trace the recent star-formation activity within the galaxy. This shows that the jet-mode AGN have a strong D$_n$(4000)  and weak Balmer absorption, which implies that these galaxies have been quiescent for more than a Gyr \citep{BRUZUAL03}. That is, the observed AGN with a lifetime of $\sim$ 100\,Myr are not immediately responsible for the  quenching of recent star formation. However, they may play a crucial role in maintaining quiescence, at least over the past 7\,Gyr.

\section{Conclusions}
\label{conclusion}

Maintenance-mode feedback from central BHs is a key element of all galaxy formation models in a cosmological context. Jet-mode AGN are the physical manifestation of this concept and a minimum requirement for the model in general is that jet-mode AGN frequently occur in galaxies devoid of significant levels of star formation. This hypothesis has thus far only been tested directly in the present-day universe but in this paper we investigate whether jet-mode galaxies at z $\sim$ 1 have been quiescent for an extended period of time.

We select the radio-loud subset of galaxies in the LEGA-C spectroscopic survey sample by matching against the newly acquired VLA 3\,GHz dataset \citep{SMOLCIC17}. We identify 58 radio-loud galaxies, most of which ($\sim 60\%$) are confirmed to be low-excitation radio AGN. Most radio sources appear point-like, but 12 sources show clear jet-like morphologies and are classified as FRI types.

The galaxies that host these jet-mode AGN have high stellar velocity dispersions of $\sigma_*>$\,175\,km\,s$^{-1}$, translating into a black-hole mass threshold of $\sim10^8$\,M$_{\odot}$ for jet-mode AGN, low specific star-formation rates ($<10^{-1}$\,Gyr$^{-1}$) and high stellar masses ($>10^{11}$\,M$_{\odot}$).
The fraction of jet-mode AGN is $\sim$ 30\% among galaxies with the highest stellar masses $\gtrsim$ 10$^{11}$\,M$_{\odot}$. Furthermore, strong 4000\AA\ breaks and weak Balmer absorption lines imply that these galaxies have been devoid of significant star-formation activity for more than $\sim 1$\,Gyr.
Assuming the jet-mode AGN share similar physical properties in a certain mass bin, and considering their life time we infer that every massive, quiescent galaxy at $z\sim 1$ will switch on a jet-mode AGN about once every Gyr.

Our findings put firmer footing on the conclusions by \citet{BEST14}, who statistically link the quiescent and radio-loud populations by comparing the evolution of their respective luminosity functions out to $z\sim 1$. It therefore seems increasingly plausible that radio AGN play a crucial role across cosmic time in keeping the halo gas around massive galaxies hot, preventing further star formation. 



\section{Acknowledgements}
We thank the anonymous referee for valuable feedback.
This project has received funding from the European Research Council (ERC) under the European Union's Horizon 2020 research and innovation programme (grant agreement 683184). Based on observations made with ESO Telescopes at the La Silla or Paranal Observatories under programme ID 194.A-2005.
V.S. acknowledges support from the European Union's Seventh Frame-work program under grant agreement 337595 (ERC Starting Grant, 'CoSMass')

\bibliographystyle{aasjournal}
\bibliography{bibtex.bib}

\end{document}